\documentclass[
    ,final            
  ]
  {aipproc}

\layoutstyle{6x9}

\begin{document}
\newcommand{\dm}{\Delta m_{15}}

\title{Type Ia supernova diversity:\\Standardizing the candles}

\classification{97.60.Bw}
\keywords      {supernovae: general}

\author{Tamara M. Davis}{
  address={Dark Cosmology Centre, Niels Bohr Institute, U. Copenhagen.  Juliane Maries Vej 30, DK-2100 Copenhagen \O, Denmark},
  email={tamarad@dark-cosmology.dk}
}

\author{J. Berian James}{
  address={Institute for Astronomy, Royal Observatory, Blackford Hill, Edinburgh, EH9 3HJ United Kingdom}
}

\author{Brian P. Schmidt}{
  address={Research School of Astronomy and Astrophysics, Australian National University, Cotter Rd, Weston ACT 2612 Australia}
}

\author{Alex G. Kim}{
  address={Physics Division, Lawrence Berkeley National Laboratory, 1 Cyclotron Road Berkeley, CA 94720 USA}
}

\begin{abstract}
Future use of type Ia supernovae for cosmology aims not only to determine the equation of state of dark energy, but also to constrain possible {\em variations} in its value.  To achieve this goal, supernovae need to become better calibrated standard candles --- not only to improve the precision of the measurement, but more importantly to gain better control over systematic uncertainties in order to ensure the accuracy of the result.  

Here we report on a project to quantify the diversity in type Ia supernovae, and to look for trends and/or sub-types that can be used to improve their calibration as standard candles.  We implement a version of principal component analysis on type Ia supernova spectra.  Although the quantity of data is not sufficient to draw any firm conclusions we show that this method holds promise for, at the very least, effectively separating peculiar supernovae.  Whether it can be further used to improve the calibration of normal type Ia's remains a project for future study. 
\end{abstract}

\maketitle


\section{Introduction}

Type Ia supernovae (SNe Ia) have proven to be excellent tools for measuring the expansion history of the universe.  Nevertheless they are not perfect standard candles and future supernova experiments (such as SNAP \cite{snap05}), which aim to constrain the time variation of the equation of state of dark energy, will be greatly enhanced if the diversity in type Ia supernovae is better understood.  This point hardly needs elaboration, and I refer the reader to \cite{james06} for a more thorough discussion of the motivation for studying SN Ia diversity.

In these proceedings we present preliminary results of an investigation into using the statistical tool Principal Component Analysis (PCA) to quantify the diversity in SN Ia spectra.  PCA, frequently used in image compression, is a method by which complicated information can be simplified by expressing it in terms of an orthogonal basis set that is derived from the data.  In the case of supernova spectra it can be thought of as deriving a set of eigenspectra, or ``principal component spectra'' (PCS), with a different set of coefficients for each supernova.  
This technique is used in \cite{francis92,suzuki06} as applied to quasar spectra, and in what follows we have adapted code originally used in \cite{francis92}.  An excellent description of PCA as applied to spectra can be found in \cite{suzuki06}, and we follow their terminology here.  An alternative attempt to use PCA on the spectra of supernovae can be found in \cite{salvo06}.

Such an analysis sees utility in a number of ways.  Firstly, it allows one to reconstruct any supernova spectrum given only a small number of coefficients, rather than the value of the spectrum at each wavelength (presuming you know the eigenspectra, of course).  In addition, once you know the distribution of each coefficient you can randomly select a set of coefficients based on these distributions and reconstruct a `typical' supernova spectrum.  This procedure could be useful for simulations in which many realistic supernova spectra are required, and for filling out the distribution of spectra when only an incomplete sample exists (as is the situation we find ourselves in).  

Secondly, and possibly most importantly, this technique could be used to identify subsets of the SN Ia population with similar characteristics.  In what follows we demonstrate this concept on a small set of previously published type Ia supernova spectra near maximum light.

\begin{figure}
  \includegraphics[height=.5\textheight]{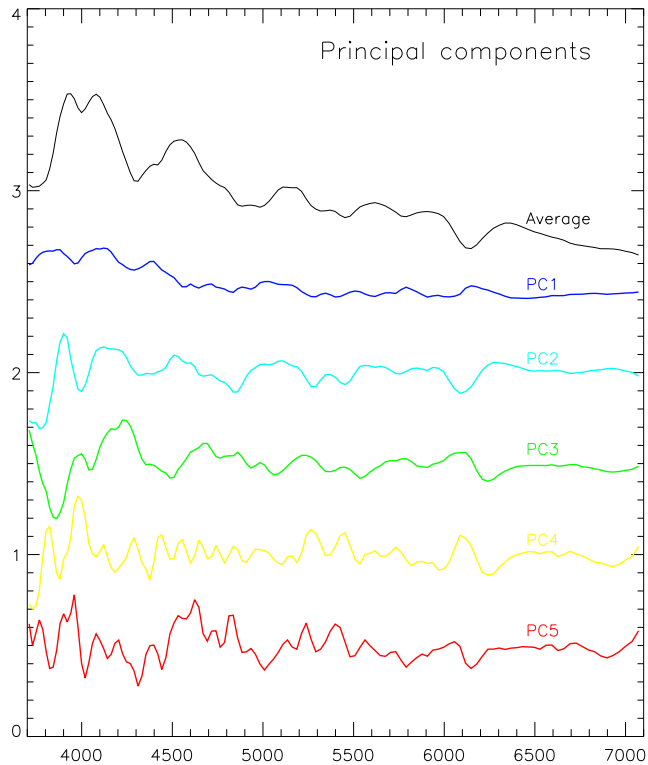}
  \label{fig:components}
  \caption{The average spectrum and the first five principal component spectra calculated for our spectral sample.  Each of the spectra in the original sample can be reconstructed from different weightings of these components. PC1 is primarily responsible for the overall shape (colour) of the spectrum.  PC2 and PC3 can be related to velocity terms that respectively red- and blue-shift the lines in the spectrum.  These first three components deal with the dominant spectral features  (representing 72\% of the variance, see Table~\ref{tab:residual}).  Higher order components become increasingly noisy.}
\end{figure}

\section{Details}

The spectral sample we used consists of 19 spectra of 10 supernovae.  These are a subset of published normal type Ia supernova spectra that were taken within two days of maximum light \cite{branch83,kirshner93,patat96,salvo01,hamuy02,branch03}.  More detail about the sample and the preparation can be found in \cite{james06}.  In summary, we performed extinction correction and warped the spectra to give the correct broadband magnitudes under synthetic photometry.  We then logarithmically binned the spectra, and combined the spectra into one average spectrum for each object. 

Since more than one spectrum was often available for each supernova we were able to use the variation in the spectra for an individual supernova to estimate the amount of non-intrinsic diversity in the sample (for example the variations due to instrumental/observational effects and the range of times about peak magnitude).  We confirmed that this intra-object dispersion was much smaller than the inter-object dispersion and consider the difference to be the size of the intrinsic diversity in SN Ia spectra.   We also confirmed that the spacing of the logarithmic bins was not significant for the final result.

\begin{figure}
  \includegraphics[height=.42\textheight]{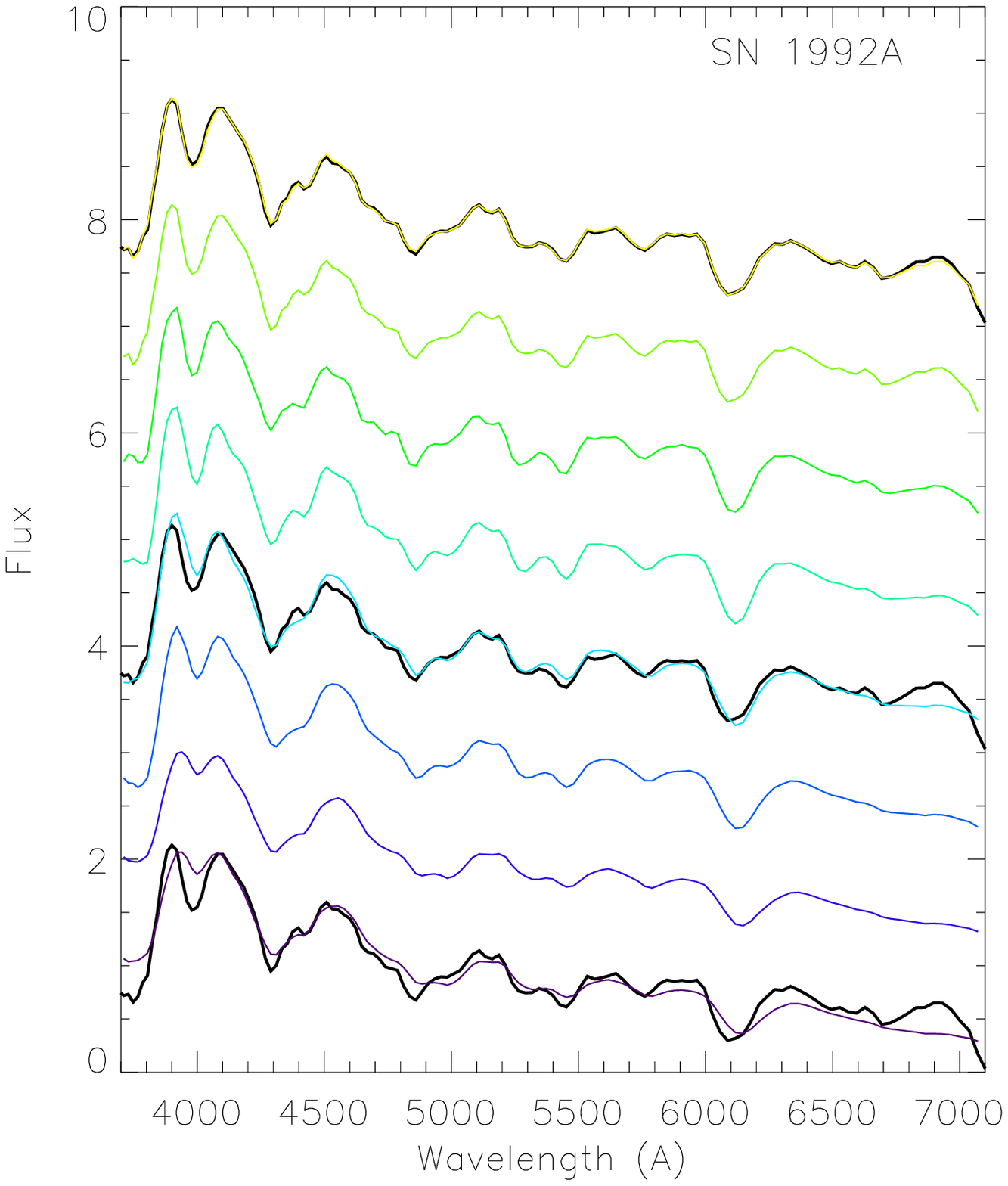}
  \includegraphics[height=.42\textheight]{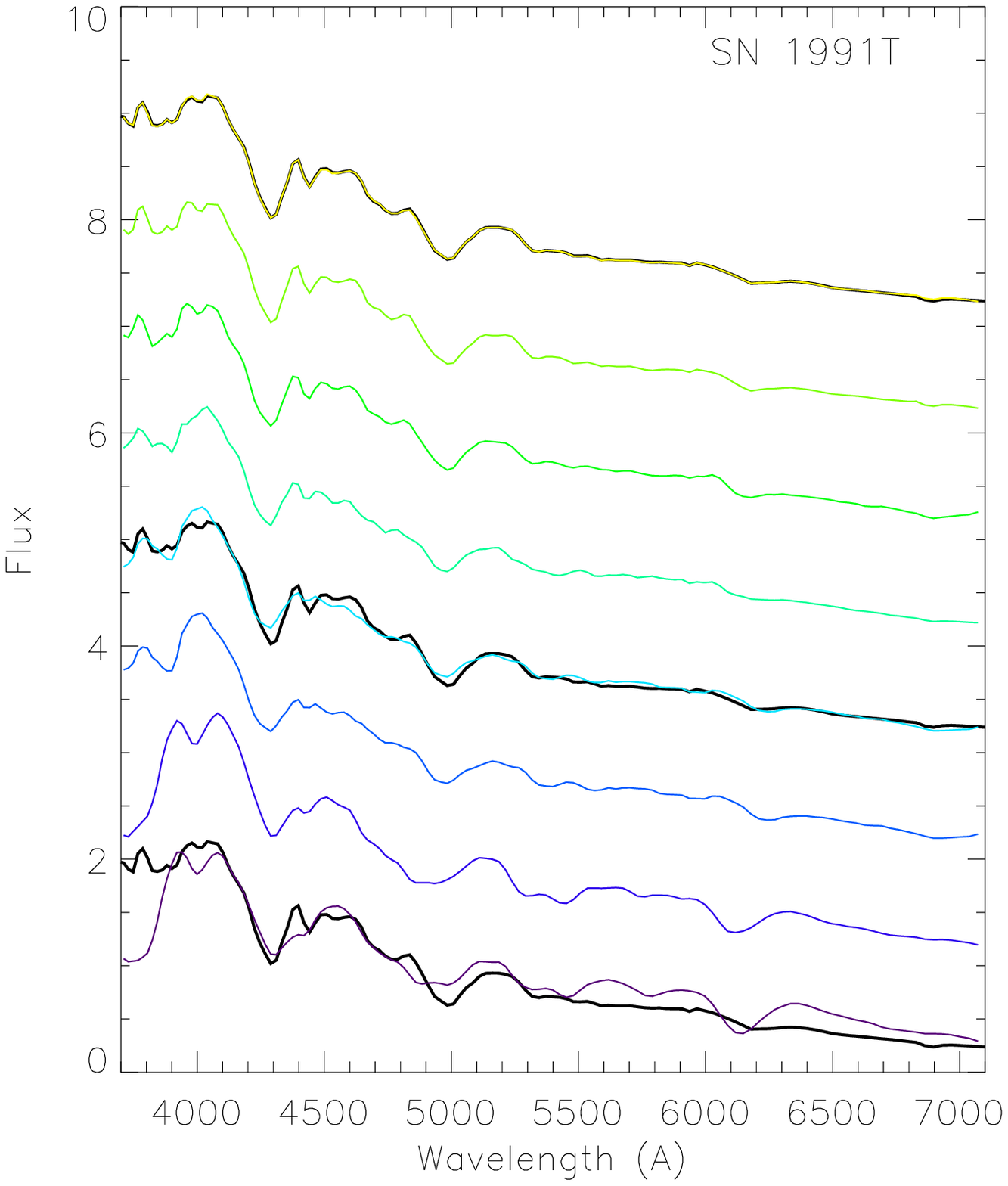}
  \label{fig:reconstruct}
  \caption{Examples of reconstructing spectra from principal components.  The lowest spectrum in each plot is the average spectrum (thin line) over-plotted with the spectrum of the supernova -- SN 1992A on the left, SN 1991T on the right. Each successive spectrum above that is the average spectrum (offset by an arbitrary amount) plus the inclusion of one additional weighted principal component.  The SN spectrum is again shown over-plotted as the thick line after 3 principal components have been added and then again after 7.  You can see that much of the variation in the spectrum has been accounted for in the first three components and by the time the 7th principal component has been included (top spectrum) the reconstructed spectrum is virtually indistinguishable from the original. This shows how two very different spectra can be reconstructed from the same principal components, weighted differently.}
\end{figure}

We then took these spectra and performed PCA.  The first five of the resulting principal component spectra are shown in Fig.~\ref{fig:components}.   In Fig.~\ref{fig:reconstruct} we demonstrate how two very different supernovae, SN 1992A and SN 1991T, can both be reconstructed by using different weightings of these principal components. 

Now that we have the components we can measure how much of the variation in the spectral sample can be accounted for by each component.  Table~\ref{tab:residual} shows the amount of variation absorbed by each component.  In this case the first four components absorb 80\% of the variation in the spectral sample, and eight components account for essentially all (98\%) of the variation.  In fact this is probably over-correcting the spectra, as not all variation in the input spectra is intrinsic to the supernovae.  The last few components in this analysis will be fitting instrumental and observational noise.  

\begin{table}[b]
\begin{tabular}{ccc}
\hline
    \tablehead{1}{c}{b}{Component\\}
  & \tablehead{1}{c}{b}{Variance\\fraction}
  & \tablehead{1}{c}{b}{Cumulative\\variance} \\
\hline
1  & 0.40 & 0.40 \\
2  & 0.17 & 0.57 \\
3  & 0.15 & 0.72 \\
4  & 0.07 & 0.80 \\
5  & 0.07 & 0.87 \\
6  & 0.05 & 0.92 \\
7  & 0.04 & 0.96 \\
8  & 0.02 & 0.98 \\
\hline
\end{tabular}
\caption{Residual variance fraction.  Amount of variation in the spectral sample that can be accounted for by each principal component.}
\label{tab:residual}
\end{table}

\begin{figure}
  \includegraphics[height=.35\textheight]{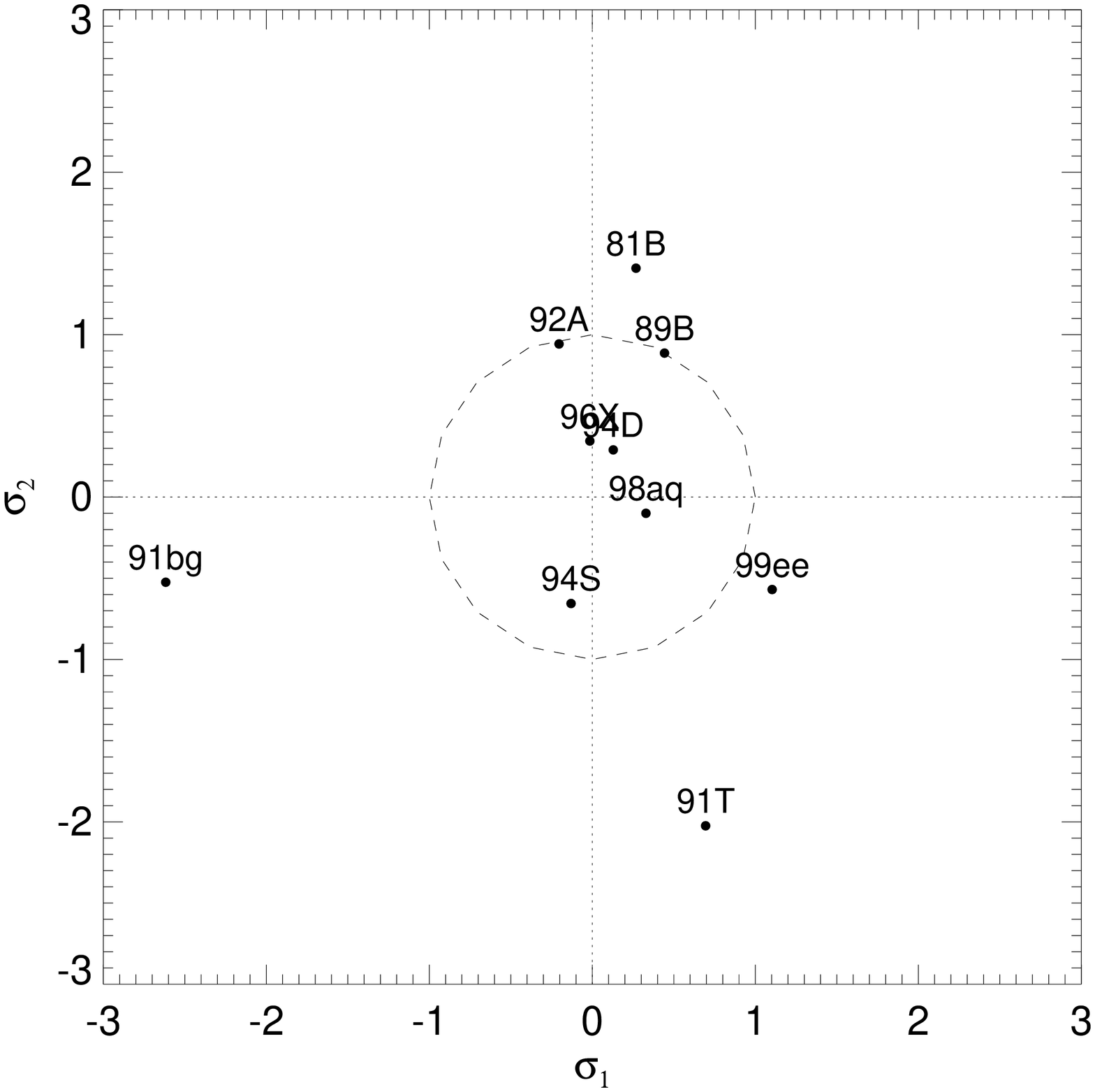}
  \includegraphics[height=.35\textheight]{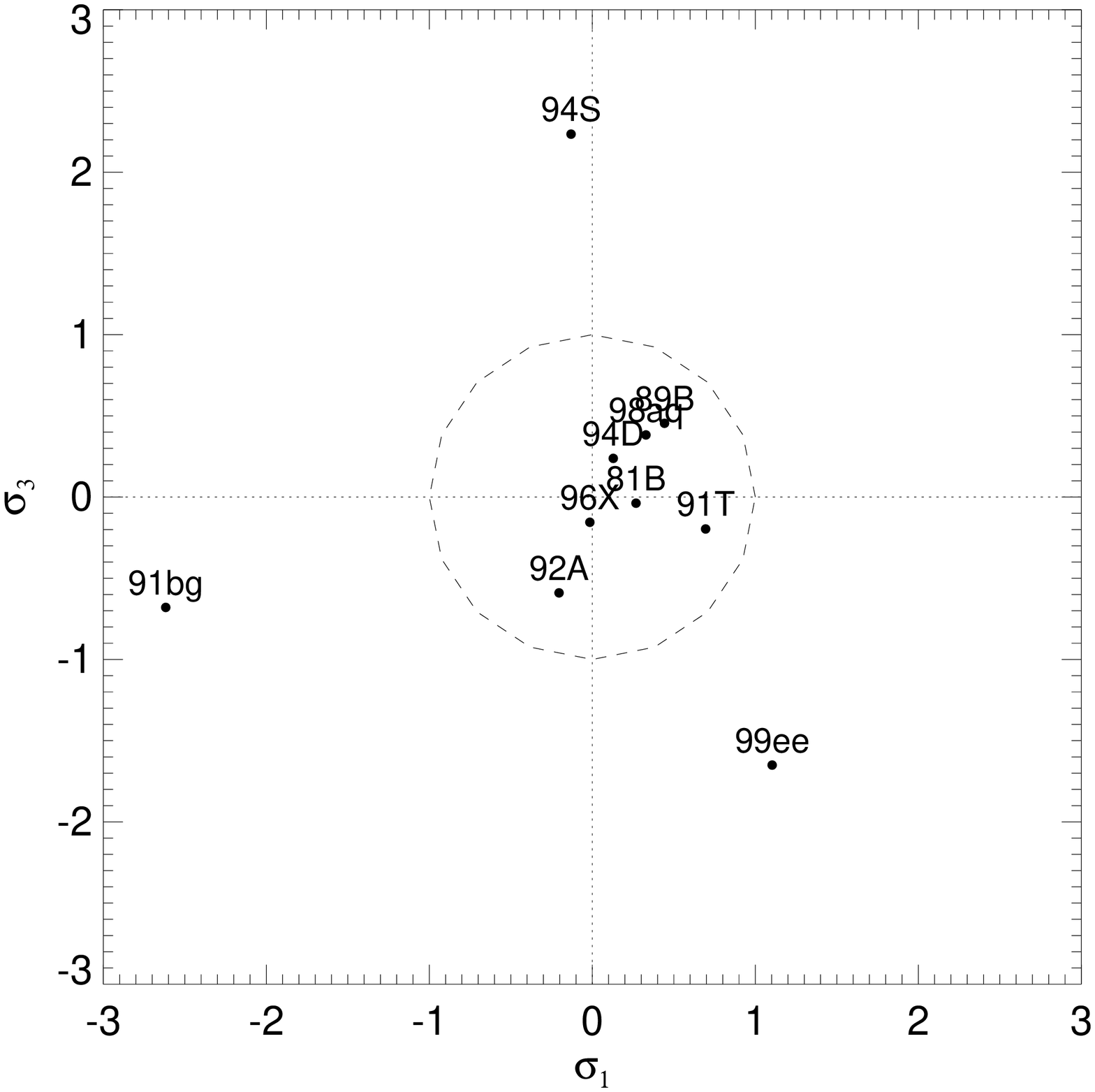}
  \label{fig:bullseye}
  \caption{Plots of the weights (coefficients) of the principal component spectra for each supernova in our sample.  Weights have been normalized to represent standard deviations.  In the left panel we see that SN 1991bg is a 2.5 $\sigma$ outlier in the first component, while SN 1991T is the furthest outlier in the second component.  The third component picks up the difference between SN 1994S and SN 1999ee, which is primarily a difference in line velocity. }
\end{figure}

The individual supernova spectra in our sample can be compared by considering their coefficients (or weights) in each of the principal component spectra.  Spectra that are similar should have similar arrays of coefficients (similar weightings of each component).  In Fig.~\ref{fig:bullseye} we plot the weight of the first component against the weight of the second (left) and third (right) for each of the supernovae in our sample.  The weights have been normalized so they represent standard deviations from the mean.  It is clear that in this diagram that SN 1991bg is peculiar in the first component, being more than 2.5 $\sigma$ from the average, while SN 1991T is somewhat peculiar in the second component.  The third component picks out the differences between SN 1994 S and SN 1999ee.  

This is a promising result, because the two SNe that are known to be peculiar, i.e. the under-luminous SN 1991bg and the over-luminous SN 1991T are picked out as the most peculiar objects in the first two principal components.  It is significant that they are peculiar in two {\em different} components, showing that this test can not only pick out peculiar supernovae, but also distinguish between the spectra of the over-luminous and under-luminous supernovae.  It will be interesting to see whether this remains a robust test with a larger spectral sample,  when more than one example from each class is used in the analysis.  

\vspace{2cm}
\section{Future work}

There is much yet to be done on this analysis.  First and foremost it needs to be applied on a much larger sample of spectra.  These are becoming available as more groups proceed with low-redshift searches and publish their results.\footnote{For example the CfA Supernova Archive has recently become publicly available and will be a great resource for this kind of study, see \url{http://www.cfa.harvard.edu/cfa/oir/Research/supernova/SNarchive.html}.}

The components of the spectra need to be related to other features of the supernovae, such as peak magnitude, $\dm$ or stretch, ejecta velocities, line ratios and equivalent widths.  Only when we have understood the relationship between these features and the different types of spectra will we have a useful tool for stamping out systematic error in our cosmology. 

Currently we have only looked at spectra near the time of peak brightness.  This can be extended to other epochs, and most powerfully to a combined multi-epoch analysis.  Once this is done we need to turn it into a tool that takes spectra that are not part of the original sample and tests where they fit in the array of diversity.  

Finally we may be able to run PCA on grids of model supernova spectra and thus relate the components to physical features such as pressure, temperature and metallicity. This could give us a much more complete understanding of the causes of diversity in, and the range of progenitors of, type Ia supernovae.

\begin{theacknowledgments}
We thank Paul Francis for the initial version of the PCA code used in this paper. TMD appreciates the support of the Villum Kann Rasmussen Fonden.  The Dark Cosmology Centre is funded by the Danish National Research Foundation.
\end{theacknowledgments}



\bibliographystyle{aipproc}   

\bibliography{tamaradavis}

\IfFileExists{\jobname.bbl}{}
 {\typeout{}
  \typeout{******************************************}
  \typeout{** Please run "bibtex \jobname" to optain}
  \typeout{** the bibliography and then re-run LaTeX}
  \typeout{** twice to fix the references!}
  \typeout{******************************************}
  \typeout{}
 }

\end{document}